THE EUROPEAN
PHYSICAL JOURNAL C

Regular Article - Theoretical Physics

# Magnetic field geometry in rotating wormhole spacetimes

Milos Ertola Urtubey[a], Daniela Pérez

Instituto Argentino de Radioastronomía (IAR, CONICET/CIC/UNLP), Villa Elisa, C.C.5, 1894 Buenos Aires, Argentina

**Abstract** If a black hole is immersed in a magnetosphere, its rotational energy can be transferred to the electromagnetic field and escape as a Poynting flux to infinity. This process of extraction of rotational energy is known as the Blandford–Znajek mechanism. It relies on the presence of both a magnetosphere and an ergosphere surrounding the black hole. Previously, we showed that rotating wormholes are also capable of emitting a Poynting ?ux in the process of accreting magnetized matter. In this work, we re-examine the Blandford–Znajek mechanism in the case of a Kerr-type wormhole. For the first time, we solve the stream equation in a rotating wormhole spacetime and derive analytical expressions for the magnetic field. We then compute the associated Poynting flux. Our results indicate that the electromagnetic flux in the wormhole spacetime is weaker than in the Kerr case, and this difference becomes more pronounced as the geometries of the two spacetimes increasingly deviate.

## 1 Introduction

The Blandford–Znajek (BZ) mechanism [1] is currently understood to refer to any process involving the extraction of rotational energy from a black hole via an electromagnetic field [2]. It has been argued that the event horizon of the black hole was one of the key elements for this process to occur. Both analytical and numerical work, however, have shown that the presence of an event horizon is not a necessary condition for the BZ mechanism to operate. Instead the ergosphere and magnetosphere appear to be the primary components responsible for the extraction of rotational energy from the compact object (see [3] and references therein). The latter implies that in an spacetime region with an ergosphere together with a magnetosphere, the BZ mechanism could take place. Such conditions could be realized, for example, in a rotating wormhole immersed in a magnetic field.

Wormholes are spacetime regions characterized by non-trivial topology. The presence of a throat and the absence of event horizons allow particles and fields to traverse between distant regions of the universe. Although wormholes remain hypothetical objects, considerable effort has been devoted to exploring their potential astrophysical signatures [4–6]. In the case of rotating wormhole solutions, various works have analyzed the motion of spinning particles in these geometries, computed the epicyclic frequencies and quasinormal oscillations in the equatorial plane, and investigated gravitational lensing [7–12].

In a recent paper [13], we show for the first time that rotating wormholes are capable of emitting a Poynting flux in the process of accreting magnetized matter. We assumed that the rotating wormhole is described by the Darmour-Solodukhin spacetime metric [14]. This Kerr-type wormhole has an ergosphere; we showed that when the ergoshere is embedded in a magnetic field, the latter supported by currents in an accretion disk, an electromagnetic flux is emitted due to the BZ mechanism. We found that for highly rotating wormholes the outgoing flux is of the same order as for a Kerr black hole.

Since our main goal was to determine whether the BZ mechanism was possible in rotating wormholes, we assumed a particular geometry for the magnetic field: initially, the magnetic field outside the ergosphere has only poloidal components. Due to the frame-drag phenomena, once the magnetic field penetrates the ergosphere, it develops a toroidal component. Although this magnetic field structure is supported by 3D GRMHD simulations [15], its precise geometry can only be determined by solving the stream equation in the given spacetime background. This is the central objective of the present work: we solve the stream equation using the Blandford–Znajek perturbative approach, extending the solution to second order in the spin parameter. This allows us

[a] e-mail: meusay@fcaglp.unlp.edu.ar (corresponding author)

Springer

to determine the magnetic field geometry within the ergoregion and to compute the associated Poynting flux. This, in turn, allow us to examine the validity of the previous model.

We find that differences between the magnetic field in the background Kerr spacetime and that in the Damour–Solodukhin wormhole spacetime arise at second order in the spin parameter, and these differences become more pronounced as the wormhole geometry deviates further from that of Kerr. Additionally, the Poynting flux is of the same order of magnitude as in our earlier estimates, thus confirming the reliability of the magnetic field configuration assumed in that model.

In the following section, we introduce the Damour–Solodukhin spacetime metric and summarize its main properties. We then provide a brief overview of the Blandford–Znajek mechanism (Sect. 3). The procedure for solving the stream equation, along with the derivation of the first- and second-order terms, is presented in Sect. 3.1. In Sect. 3.2, we display plots of the magnetic field lines near the wormhole's throat. Subsequently, we estimate the Poynting flux and compare our results with those obtained in our previous work. We close the paper with some conclusions and perspectives.

## 2 Damour–Solodukhin rotating wormhole spacetime metric

The Kerr-type wormhole metric derived in [14] is a generalization of the static, non-rotating wormhole metric derived by Damour and Solodukhin in their seminal paper [16] (from here onward we will to it as RDSW (Rotating Damour-Solodukhin Wormhole). This metric is characterized by three parameters: the wormhole mass $M$, the spin $a \equiv J/M$, and the deformation parameter $\lambda$. In Boyer-Lindquist coordinates $(t, r, \theta, \phi)$ the line element takes the form

$$ds^2 = -\left(1 - \frac{2GMr}{c^2\Sigma}\right)c^2 dt^2 - \frac{4GMar\sin\theta^2}{c\Sigma}dtd\phi$$
$$+ \frac{\Sigma}{\hat{\Delta}}dr^2 + \Sigma d\theta^2 + \left(r^2 + a^2 + \frac{2GMa^2 r\sin\theta^2}{c^2\Sigma}\right)$$
$$\times \sin^2\theta d\phi^2. \tag{1}$$

Here, $\Sigma$ and $\hat{\Delta}$ represent auxiliary functions, $\Sigma \equiv r^2 + a^2\cos^2\theta$, $\hat{\Delta} \equiv r^2 - 2GM/c^2(1+\lambda^2)r + a^2$. The radial coordinate $r$ is defined over the interval $r_+ \leq r < \infty$, where $r_+$ is the largest solution of the equation $\hat{\Delta} = 0$. In the limit $\lambda \to 0$, the Kerr black hole metric is recovered.

The Damour-Solodukhin wormhole was introduced as a phenomenological deformation of the Schwarzschild solution, rather than derived from a fundamental action. The same approach was taken for the case of the RDSW. Nevertheless, if one substitutes the RDSW metric into Einstein's equations, the resulting geometry can be interpreted as a solution with an effective stress–energy tensor. This effective matter content necessarily violates the energy conditions, as expected for traversable wormholes (the violation of the energy conditions was shown in [17]). While no simple local Lagrangian is known to yield exactly the RDSW metric, in principle such an energy–momentum tensor could be modeled by suitable exotic fields.

Similarly to the Kerr metric, the line element (1) has two apparent singularities determined by the values of $r$ and $\theta$ such that $\Sigma = 0$ and $\hat{\Delta} = 0$, these singularities representing the two boundary surfaces of the wormhole, the ergosphere, and the throat.

In Boyer–Lindquist coordinates, the radial coordinate of the ergosphere and the throat take the form

$$r_{S\pm} = \frac{GM}{c^2} \pm \sqrt{\left(\frac{GM}{c^2}\right)^2 - a^2\cos^2\theta}, \tag{2}$$

$$r_{\pm} = \frac{GM}{c^2}(1+\lambda^2) \pm \sqrt{\left(\frac{GM}{c^2}\right)^2(1+\lambda^2)^2 - a^2}, \tag{3}$$

respectively.

In Fig. 1 we show a representation of the wormhole ergosurface and throat for fixed values of $a$ and $M$, varying the parameter $\lambda$. In the limit $\lambda = 0$, we recover the Kerr line element, and the surfaces coincide with the black hole's ergosphere and event horizon. As the deformation parameter increases, the throat becomes larger, while the ergosphere remains unchanged. Consequently, the ergosphere does not fully enclose the throat.[1]

## 3 Blandford–Znajek mechanism and stream equation

The Blandford–Znajek mechanism is defined as the process of extracting rotational energy from a compact object via an electromagnetic field. Since the event horizon is causally disconnected from the outflow, it is not required for the mechanism to occur. Instead, the essential elements are identified as the ergosphere and a magnetic field, allowing the mechanism to occur not only in black holes, but also in rotating wormholes.

Let us consider a magnetosphere resulting from the flow of currents in an accretion disk across the equator of the compact object in the force-free regime. As a result, the magnetosphere is such that the Lorentz force is zero and the astrophysical plasma pressure is dominated by the magnetic pressure. Under these conditions, the dynamics of the magnetosphere are described by Maxwell's equations.

$$\nabla_\mu F^{\mu\nu} = j^\nu, \quad \nabla_{[\rho}F_{\mu\nu]} = 0, \tag{4}$$

---

[1] For a deeper analysis of this spacetime, see [13].

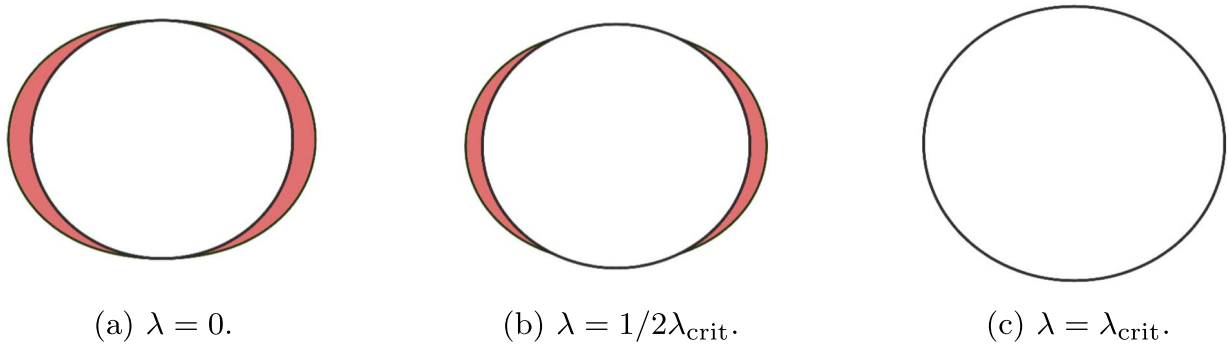

(a) $\lambda = 0$.  (b) $\lambda = 1/2\lambda_{\text{crit}}$.  (c) $\lambda = \lambda_{\text{crit}}$.

**Fig. 1** Schematic diagram of the throat and ergoregion of the RDSW for different values of the deformation parameter. The throat is the white region with black rim and the ergoregion is the red section with black rim

under the force-free

$$F_{\mu\nu}J^\nu = 0, \qquad (5)$$

and MHD ideal condition

$$F^{*\mu\nu}F_{\mu\nu} = 0. \qquad (6)$$

Based on these considerations, the conservation of the energy-momentum tensor, given by the equation $\nabla_\mu T^\mu_\nu = 0$, implies that the electromagnetic energy flow in the $\nu$ direction, represented by the quantity $T^\nu_t$, is conserved. In particular, an electromagnetic energy flow in the radial direction can be defined as follows [1,18,19]

$$\begin{aligned}\mathcal{E}^r &= -T^r_t = \left(F_{t\theta}F_{\theta\phi}g^{r\phi} - F_{r\theta}F_{t\theta}g^{rr} - F_{t\theta}^2 g^{tr}\right)g^{\theta\theta},\\ &= -c^2\omega(r,\theta)B^r(r,\theta)B^\phi(r,\theta)\Delta\sin^2\theta\end{aligned} \qquad (7)$$

To compute this quantity, the geometry of the magnetic field must be known.

To determine this geometry in the background spacetime, the following assumptions are typically made: given that the spacetime is stationary and axisymmetric, it is assumed that the magnetosphere exhibits the same symmetries. Consequently, a gauge is chosen such that the four-vector potential is independent of the temporal and azimuthal coordinates: Thus, we have $\partial_t A_\mu = \partial_\phi A_\mu = 0$.

Next, three functions are defined:

1. $\Psi(r,\theta) \equiv A_\phi(r,\theta)$, which represents the magnetic flux through a circular loop of radius $r\sin\theta$ surrounding the polar axis of the compact object.
2. The angular velocity of the magnetic field lines:

$$\Omega(r,\theta) = -\frac{\partial_r A_t}{\partial_r \Psi} = -\frac{\partial_\theta A_t}{\partial_\theta \Psi}. \qquad (8)$$

3. The poloidal current:

$$I = -\sqrt{-\frac{g_T}{g_P}} F_{r\theta}. \qquad (9)$$

where $g_T$ and $g_P$ are the determinants

$$\begin{aligned}g_T &= g_{tt}g_{\phi\phi} - g_{t\phi}^2 = -\Delta\sin^2\theta, \quad g_P = g_{rr}g_{\theta\theta}\\ &= \frac{\Sigma^2}{\hat\Delta}, \quad g = g_T g_P = -\Sigma^2\frac{\Delta}{\hat\Delta}\sin^2\theta.\end{aligned} \qquad (10)$$

Note that $\Omega$ and $I$ are functions that depend only on the magnetic flux, so $\Omega = \Omega(\Psi)$ and $I = I(\Psi)$ [20].

Given these definitions and all previous assumptions, from Eqs. (4), (5) and (6) a partial differential equation for the magnetic flux $\Psi(r,\theta)$ can be derived, known as the stream equation [20–22]:

$$\Omega\partial_\rho\left(\sqrt{-g}F^{t\rho}\right) - \partial_\rho\left(\sqrt{-g}F^{\phi\rho}\right) - F_{r\theta}\frac{dI}{d\Psi} = 0. \qquad (11)$$

A common approach to obtaining analytical solutions to this equation is to apply a perturbative method around the split-monopole solution, expanding in terms of the spin parameter of the compact object[2] [1,20,22,25,26].

In the following section we solve the stream equation in the RDSW background spacetime, using a perturbative method up to second order on the spin parameter $a$. To the best of our knowledge, this has never been done before for the case of a rotating wormhole.[3]

### 3.1 Perturbative solution of the stream equation

For an axisymmetric spacetime, the stream equation takes the form

$$\eta_\mu\partial_\nu\left(\eta^\mu\sqrt{-g}g^{\nu\rho}\partial_\rho\Psi\right) = F_{r\theta}\frac{dI}{d\Psi}, \qquad (12)$$

---
[2] The BZ mechanism has also been studied in alternative theories of gravity, where the same approach was adopted to solve the stream equation (see for instance [23,24]).

[3] The properties of electric and magnetic fields for static spherically symmetric wormholes were analyzed in [27,28].

where $\eta_\mu = \delta_\mu^\phi - \Omega \delta_\mu^t$, $F_{r\theta} = -I\sqrt{-\frac{g_P}{g_T}}$, $g_P = g_{rr}g_{\theta\theta} = \frac{\Sigma^2}{\hat{\Delta}}$, $g_T = g_{tt}g_{\phi\phi} - g_{t\phi}^2 = -c^2\Delta\sin^2\theta$, and the metric determinant $g = g_T g_P = -c^2\Sigma^2\sin^2\theta\Delta/\hat{\Delta}$.

Now, we will consider a perturbative expansion up to second order over the normalized spin parameter $\alpha = 2a/r_0$, with $r_0 = 2GM/c^2$. Under this assumption, all relevant expressions take the following form:

$$\eta_\mu = \delta_\mu^\phi - (\alpha\omega_1(r,\theta) + \alpha^2\omega_2(r,\theta)) + \mathcal{O}(\alpha^3), \quad (13)$$
$$\Psi(r,\theta) = \Psi_0(r,\theta) + \alpha\Psi_1(r,\theta) + \alpha^2\Psi_2(r,\theta) + \mathcal{O}(\alpha^3), \quad (14)$$
$$\Omega(r,\theta) = \omega_0(r,\theta) + \alpha\omega_1(r,\theta) + \alpha^2\omega_2(r,\theta) + \mathcal{O}(\alpha^3), \quad (15)$$
$$I(r,\theta) = i_0(r,\theta) + \alpha i_1(r,\theta) + \alpha^2 i_2(r,\theta) + \mathcal{O}(\alpha^3), \quad (16)$$

with $\Psi_i(r,\theta)$, $\omega_i(r,\theta)$ and $i_i(r,\theta)$ the field functions at the $i$-subleading order.

As we mentioned previously, our seed solution is the split-monopole configuration where we follow [20]

$$\Psi_0(r,\theta) = \mathcal{K}(1-\cos\theta), \qquad \omega_0(r,\theta) = 0,$$
$$i_0(r,\theta) = 0, \quad (17)$$

being $\mathcal{K}$ a constant that has the proper units for the magnetic flux. Substituting these expressions in the stream equation, we obtain

$$A + B\alpha + C\alpha^2 = 0. \quad (18)$$

Being a series expansion of an equation, for it to be satisfied, we need that each term must also be null. Next, we compute explicitly $A$, $B$ and $C$:

$$A = \partial_r\left(\frac{\sqrt{(c^2r - 2GM)(c^2r - 2GM(1+\lambda^2))}}{c^2 r \sin\theta}\partial_r \Psi_0\right)$$
$$+ \partial_\theta\left(\sqrt{\frac{(c^2r - 2GM)}{(c^2r - 2GM(1+\lambda^2))}}\frac{1}{r^2\sin\theta}\partial_\theta\Psi_0\right),$$
$$= \partial_r\left(\frac{\sqrt{(c^2r - 2GM)(c^2r - 2GM(1+\lambda^2))}}{c^2 r \sin\theta}\partial_r(\mathcal{K}(1-\cos\theta))\right)$$
$$+ \partial_\theta\left(\sqrt{\frac{(c^2r - 2GM)}{(c^2r - 2GM(1+\lambda^2))}}\frac{\partial_\theta \mathcal{K}(1-\cos\theta)}{r^2\sin\theta}\right),$$
$$= \partial_\theta\left(\sqrt{\frac{(c^2r - 2GM)}{(c^2r - 2GM(1+\lambda^2))}}\frac{1}{r^2\sin\theta}\mathcal{K}\sin\theta\right) = 0. \quad (19)$$

So, term $A = 0$ for our choice of the seed solution.

### 3.1.1 First order term

The linear term has the form

$$B = \partial_r\left(\frac{\sqrt{(c^2r - 2GM)(c^2r - 2GM(1+\lambda^2))}}{c^2 r \sin\theta}\partial_r \Psi_1\right)$$
$$+ \partial_\theta\left(\sqrt{\frac{(c^2r - 2GM)}{(c^2r - 2GM(1+\lambda^2))}}\frac{\partial_\theta \Psi_1}{r^2\sin\theta}\right).$$

The equation $B = 0$ is a partial differential equation for the unknown function $\Psi_1(r,\theta)$. We aim for a solution using the method of separable variables, where we define

$$\Psi_1(r,\theta) = R_1(r)\Theta_1(\theta).$$

Then, we transform the partial differential equation for $\Psi_1(r,\theta)$ into a system of second-order linear differential equations (ODE) for $R_1(r)$ and $\Theta_1(\theta)$

$$\frac{d}{dr}\left(\frac{1}{c^2 r}\sqrt{(c^2r - 2GM)(c^2r - 2GM(1+\lambda^2))}\frac{dR_1(r)}{dr}\right)$$
$$= \mathcal{D}\sqrt{\frac{(c^2r - 2GM)}{(c^2r - 2GM(1+\lambda^2))}}\frac{R_1(r)}{r^2},$$
$$\frac{d}{d\theta}\left(\frac{1}{\sin\theta}\frac{d\Theta_1(\theta)}{d\theta}\right)$$
$$= -\frac{\mathcal{D}}{\sin\theta}\Theta_1(\theta).$$

Considering $\mathcal{D} = l(l+1)$, the angular solutions are given in terms of the hypergeometric functions,

$$\Theta_{1,2k-1}(\theta) = {}_2F_1\left(-k, k-\tfrac{1}{2}; \tfrac{1}{2}; \cos^2\theta\right), \quad (20)$$
$$\Theta_{1,2k}(\theta) = {}_2F_1\left(-k, k+\tfrac{1}{2}; \tfrac{3}{2}; \cos^2\theta\right), \quad (21)$$

where $k$ is a positive integer.

To solve the radial ODE, we first perform the variable change $x = rc^2/(2GM(1+\lambda^2))$. The ODE follows

$$R''(x) + p(x)R'(x) + q(x)R(x) = 0, \quad (22)$$

where

$$p(x) = \frac{b(\frac{b+1}{2b}x - 1)}{x(x-1)(x-b)}, \quad (23)$$
$$q(x) = \frac{-l(l+1)}{x(x-1)}, \quad (24)$$
$$b = \frac{1}{1+\lambda^2}. \quad (25)$$

This is a second-order differential equation with 4 regular singularities, $0$, $b$, $1$, $\infty$, and hence can be represented as a Heun equation,

$$R''(x) + \left(\frac{\gamma}{x} + \frac{\delta}{x-1} + \frac{\epsilon}{x-a}\right)R'(x)$$
$$+ \frac{\alpha\beta x - q}{x(x-1)(x-a)}R(x) = 0, \quad (26)$$
$$\epsilon = 1 + \alpha + \beta - \gamma - \delta, \quad (27)$$

with $q = -l(l+1)b$, $\alpha = l$, $\beta = -(l+1)$, $\gamma = -1$, $\delta = \tfrac{1}{2}$, $\epsilon = \tfrac{1}{2}$.

By a similar analysis to the one done in [20,22], we find that the only solution for this ODE that is both regular at the throat ($x = 1$) and at infinity is the null solution,

$$R_1(r) = 0.$$

Then, the linear term of the magnetic flux is also null, $\Psi_1(r,\theta) = 0$, as is the case for the Kerr black hole metric.

Next, we follow an approach similar to that of [22] to determine the magnetic frequency $\omega_1(r,\theta)$ and the poloidal current $i_1(r,\theta)$. The integrability conditions,

$$\partial_r \Psi \partial_\theta \Omega = \partial_\theta \Psi \partial_r \Omega,$$
$$\partial_r \Psi \partial_\theta I = \partial_\theta \Psi \partial_r I,$$

yield:

$$\omega_1 = \omega_1(\theta), \quad i_1 = i_1(\theta). \tag{28}$$

It is well known that the light surfaces, i.e., surfaces where the world line of a particle is null, have a particular physical significance in black hole magnetospheres [29]. These surfaces, referred to as the Inner Light Surface (ILS) and the Outer Light Surface (OLS), correspond to locations where the velocity of an observer co-rotating with the magnetic field lines becomes null. They are determined by the solutions of the equation:

$$g_{tt} + 2\Omega g_{t\phi} + \Omega^2 g_{\phi\phi} = 0. \tag{29}$$

The physical relevance of these surfaces can be seen by studying how charged particles behave inside the ILS (this surface is always outside the black hole horizon) and outside the OLS: the charged particles inside the ILS are forced to go towards the black hole horizon, while outside the OLS, they are forced to move away from it.

Note that Eq. (29) (for further details we refer to [20]) is the same for both the RDSW and Kerr spacetime. The first-order corrections to these surfaces in the RDSW spacetime coincide with those found in the Kerr black hole case.

$$r_{\text{ILS}} = r_0 - r_0 \frac{1}{4} \left( \cos^2\theta + 4r_0 \sin^2\theta \omega_1(\theta) \right.$$
$$\left. -4r_0^2 \sin^2\theta \omega_1^2(\theta) \right) \alpha^2 + \mathcal{O}(\alpha^3),$$
$$r_{\text{OLS}} = \frac{1}{\omega_1(\theta)\sin\theta}\frac{1}{\alpha} + \mathcal{O}(1).$$

Additionally, the light surfaces are such that

$$g^{\mu\nu}\eta_\mu\eta_\nu\Big|_{r_{\text{ILS,OLS}}} = 0. \tag{30}$$

The latter implies that (12) can be written as

$$\sqrt{-g}g^{rr}\eta_\mu\partial_r\eta^\mu\partial_r\Psi + \sqrt{-g}g^{\theta\theta}\eta_\mu\partial_\theta\eta^\mu\partial_\theta\Psi$$
$$+ I\sqrt{-\frac{g_P}{g_T}}\frac{dI}{d\Psi}\bigg|_{r_{\text{ILS}},r_{\text{OLS}}} = 0. \tag{31}$$

From Eqs. (12) and (31), we obtain a linear equation system for $\omega_1(\theta)$ and $i_1(\theta)$,

$$\partial_\theta i_1(\theta)^2 = \partial_\theta \left( \mathcal{K}\frac{1}{2r_0}\left(1 - \frac{2r_0}{c}\omega_1(\theta)\right)\sin^2\theta \right)^2, \tag{32}$$

$$\partial_\theta i_1(\theta)^2 = \partial_\theta \left( \mathcal{K}\frac{\omega_1(\theta)}{c}\sin^2\theta \right)^2, \tag{33}$$

where $\mathcal{K}$ is a proportionality constant given by,

$$\mathcal{K} = \frac{B_0}{c}\left(\frac{GM}{c^2}\right)^3. \tag{34}$$

The solutions of the system of equations by (32) (33) are

$$\omega_1(\theta) = \frac{c^3}{8GM},$$
$$i_1(\theta) = \mathcal{K}\omega_1(\theta)\sin\theta. \tag{35}$$

These are the same solutions as the ones derived for a Kerr black hole [20].

### 3.1.2 Second order term

We derive the quadratic term of the stream equation by substituting the expressions for $\omega_1(\theta)$ and $i_1(\theta)$ into 12, obtaining the following second-order PDE:

$$\partial_r f_1(r,\theta) + \partial_\theta f_2(r,\theta) =$$
$$-\frac{2GM(c^2r + 2GM)}{2c^4r^4}\sqrt{\frac{(c^2r - 2GM)}{(c^2r - 2GM(1+\lambda^2))}}\frac{\Theta_2(\theta)}{\sin\theta}, \tag{36}$$

$$\Theta_2(\theta) = 3\cos\theta\sin^2\theta, \tag{37}$$

where

$$f_1(r,\theta) = \frac{1}{c^2r}\frac{\sqrt{(c^2r - 2GM)(c^2r - 2GM(1+\lambda^2))}}{\sin\theta}\partial_r\Psi_2, \tag{38}$$

$$f_2(r,\theta) = \sqrt{\frac{(c^2r - 2GM)}{(c^2r - 2GM(1+\lambda^2))}}\frac{1}{r^2\sin\theta}\partial_\theta\Psi_2. \tag{39}$$

Proposing the solution

$$\Psi_2(r,\theta) = R_2(r)\Theta_2(\Theta), \tag{40}$$

the PDE reduces to the ODE,

$$\partial_r\left(\frac{1}{c^2r}\sqrt{(c^2r - 2GM)(c^2r - 2GM(1+\lambda^2))}\partial_r R_2(r)\right)$$
$$+ \sqrt{\frac{(c^2r - 2GM)}{(c^2r - 2GM(1+\lambda^2)}} \tag{41}$$
$$\left(-\frac{6}{r^2}R_2(r) + \frac{2GM(c^2r + 2GM)}{2c^4r^4}\right) = 0.$$

Equation (41) is a non-homogeneous ordinary differential equation (ODE), and its solution can therefore be obtained from the general solution of the homogeneous ODE, $f_{\text{hom}}(r)$, and the particular solution of the non-homogeneous equation, $f_{\text{nohom}}(r)$,

$$R_2(r) = f_{\text{hom}}(r) + f_{\text{nohom}}(r),$$

with

$$\partial_r \left( \frac{1}{c^2 r} \sqrt{(c^2 r - 2GM)(c^2 r - 2GM(1 + \lambda^2))} \partial_r f_{\text{hom}}(r) \right)$$
$$- \frac{6}{r^2} \sqrt{\frac{(c^2 r - 2GM)}{(c^2 r - 2GM(1 + \lambda^2))}} f_{\text{hom}}(r) = 0,$$

and

$$\partial_r \left( \frac{1}{c^2 r} \sqrt{(c^2 r - 2GM)(c^2 r - 2GM(1 + \lambda^2))} \partial_r f_{\text{nohom}}(r) \right)$$
$$- \sqrt{\frac{(c^2 r - 2GM)}{(c^2 r - 2GM(1 + \lambda^2))}} \frac{6}{r^2} f_{\text{nohom}}(r) =$$
$$- \sqrt{\frac{(c^2 r - 2GM)}{(c^2 r - 2GM(1 + \lambda^2))}} \frac{2GM(c^2 r + 2GM)}{2c^4 r^4}.$$

The homogeneous ODE has the same structure as Eq. (12). Therefore, the only solution that remains well-behaved both at the throat and at infinity is the trivial one, $f_{\text{hom}}(r) = 0$. Consequently, it suffices to compute only the particular solution, $f_{\text{nohom}}(r)$.

Defining $r_0 = 2GM/c^2$, $\epsilon = 1 + \lambda^2$, and introducing the change of variable $r = \epsilon r_0 x$, we find that the resulting equation is the same Heun equation encountered at first order, but with $l = 2$ and the addition of a non-homogeneous term. For convenience, we relabel the function $f_{\text{nohom}}(x)$ as $R(x)$:

$$R''(x) + \left( -\frac{1}{x} + \frac{1}{2\left(x - \frac{1}{\epsilon}\right)} + \frac{1}{2(x-1)} \right) R'(x)$$
$$- \frac{6\left(x - \frac{1}{\epsilon}\right)}{(x-1)x\left(x - \frac{1}{\epsilon}\right)} R(x)$$
$$= -\frac{\left(x + \frac{1}{\epsilon}\right)}{2\epsilon x^3 (x-1)}, \tag{42}$$

In the limit $\epsilon \to 1$ ($\lambda \to 0$), the latter reduces to

$$R''(x) + \left( -\frac{1}{x} + \frac{1}{x-1} \right) R'(x) - \frac{6}{(x-1)x} R(x)$$
$$= -\frac{x+1}{2x^3(x-1)}, \tag{43}$$

which is the known ODE in the Kerr spacetime background [22].

We were unable to find an analytical solution to Eq. (42). Therefore, we employ a numerical approach, computing solutions for the spin parameter $\alpha = 0.5$ and for the following values of the deformation parameter[4]:

$$\lambda = 0, \quad \lambda = \frac{\alpha\sqrt{16 + 9\alpha^2}}{8\sqrt{16 - 3\alpha^2}}, \quad \lambda = \frac{\alpha\sqrt{16 + 9\alpha^2}}{4\sqrt{16 - 3\alpha^2}}.$$

The solutions were obtained under the condition that they remain regular both at the throat and at spatial infinity. The corresponding results are shown in Fig. 2. We see that at large distances from the event horizon or throat, the Kerr solution dominates over the wormhole solutions. In contrast, near the throat, the wormhole solutions deviate more significantly, with the value of the radial function at the throat or event horizon increasing with the deformation parameter.

---

[4] We discuss the motivation for these specific choices of the deformation parameter below.

Given the the quadratic correction $\Psi_2$ for the magnetic flux, we now compute the functions $\omega_2(r, \theta)$ and $i_2(r, \theta)$. Once again, due to the integrability conditions (28), both functions have a dependency only on the poloidal variable, $\omega_2 = \omega_2(\theta)$, $i_2 = i_2(\theta)$.

Once again, as Eq. (29) is the same for the spacetime metric considered in this study and for the Kerr metric, the corrections for these surfaces in the RDSW spacetime coincide with the ones of a Kerr black hole,

$$r_{\text{ILS}} = r_0 - r_0 \frac{1}{4} \left( \cos^2 \theta + 4r_0 \sin^2 \theta \omega_1(\theta) - 4r_0^2 \sin^2 \theta \omega_1^2(\theta) \right) \alpha^2$$
$$- \frac{1}{2} r_0^2 \omega_2(\theta) \sin^2 \theta \alpha^3 + \mathcal{O}(\alpha^4), \tag{44}$$
$$r_{\text{OLS}} = \frac{1}{\omega_1(\theta) \sin \theta} \frac{1}{\alpha} - \frac{r_0}{2 \sin \theta} (\sin \theta + 32 r_0 \omega_2(\theta)) + \mathcal{O}(\alpha).$$

Inputting these in Equations (12),(31), the corrections for the magnetic frequency and poloidal current must be such that

$$i_2 + \frac{1}{2} \tan \theta \partial_\theta i_2(\theta)$$
$$= K \left( -2\omega_2(\theta) \sin^2(\theta) - \frac{1}{2} \sin^2 \theta \tan \theta \partial_\theta \omega_2(\theta) \right),$$
$$i_2 + \frac{1}{2} \tan \theta \partial_\theta i_2(\theta)$$
$$= K \left( 2\omega_2(\theta) \sin^2(\theta) + \frac{1}{2} \sin^2 \theta \tan \theta \partial_\theta \omega_2(\theta) \right). \tag{45}$$

These two equations are the same as the ones obtained for the Kerr metric, so they share the same solutions,

$$\omega_2(\theta) = 0, \quad i_2(\theta) = 0. \tag{46}$$

Let us recall that to find the correction terms for the magnetic frequency $\omega(r, \theta)$ and $i(r, \theta)$ we used the light surfaces $r_{\text{ILS}}$ and $r_{\text{OLS}}$. In Kerr spacetime, the ILS surface remains outside the event horizon. In the case of the RDS wormhole the radius of the throat increases with the deformation parameter (the radius of the ILS only depends on the angle $\theta$ and the functions $\omega(r, \theta)$ and $i(r, \theta)$). Hence, there is a maximum value for $\lambda$, denoted $\lambda_{\max}$, such that the ILS coincides with the throat. For $\lambda > \lambda_{\max}$ there is no ILS surface in the RRDSW spacetime.

This upper limit can be determined by considering that the throat coincides with the maximum dimension the ILS surface can take

$$r_{\text{ILS}}(\theta) \leq r_{\text{ILS}}\left(\frac{\pi}{2}\right) = r_+, \quad \forall \theta.$$

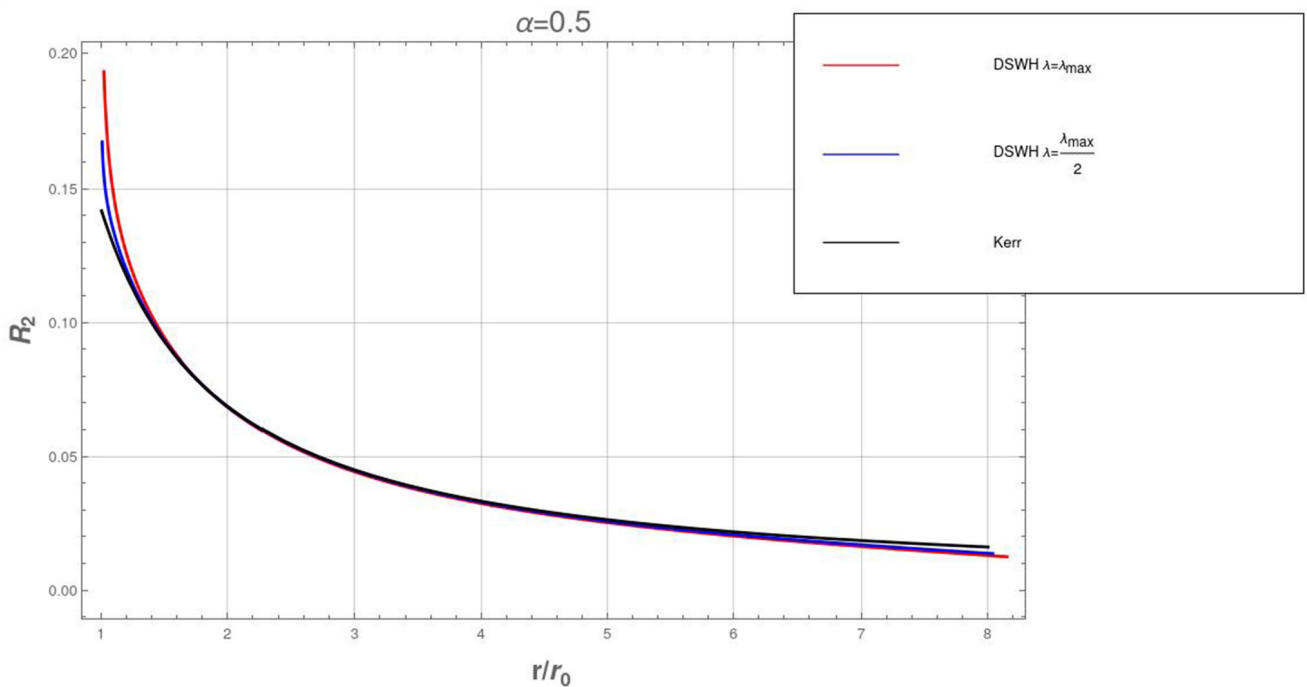

**Fig. 2** Radial solution for the stream equation with the choice $\alpha = 0.5$

We conclude that the range of values for the deformation parameter that we can adopt is $0 \leq \lambda \leq \lambda_{\max}$, with this upper limit being

$$\lambda_{\max} = \frac{\alpha\sqrt{16 + 9\alpha^2}}{4\sqrt{16 - 3\alpha^2}}. \tag{47}$$

### 3.2 Magnetic field lines

Having solved the stream equation, we can now derive the magnetic field components in Boyer–Lindquist coordinates, $B^r(r,\theta)$, $B^\theta(r,\theta)$, and $B^\phi(r,\theta)$. These components can be expressed in terms of the functions $\Psi$, $\Omega$, and $I$:

$$\begin{aligned}
B^r(r,\theta) &= \frac{1}{\sqrt{-g}} F_{\theta\phi} = \frac{1}{\sqrt{-g}} \partial_\theta \Psi \\
&= \frac{1}{\sqrt{-g}} \partial_\theta (K((1-\cos\theta) + \alpha^2 R_2(r)\Theta_2(\theta))), \\
B^\theta(r,\theta) &= -\frac{1}{\sqrt{-g}} F_{r\phi} = -\frac{1}{\sqrt{-g}} \partial_r \Psi \\
&= -\frac{1}{\sqrt{-g}} \partial_r (K((1-\cos\theta) + \alpha^2 R_2(r)\Theta_2(\theta))), \\
B^\phi(r,\theta) &= \frac{1}{\sqrt{-g}} F_{r\theta} = -\frac{I}{g_T} = -\frac{\alpha i_1(\theta)}{g_T}.
\end{aligned} \tag{48}$$

To visualize the magnetic field geometry, we express the magnetic field components in Cartesian coordinates (the procedure is detailed in Appendix 6). In Fig. 3, we present the magnetic field lines along with the wormhole's throat (left) and the black hole's event horizon (right), represented by the central sphere. The upper panel shows the magnetic field lines as viewed from the equatorial plane of the wormhole ($\theta = \pi/2$), while the lower panel displays the field lines as seen from the polar direction ($\theta = 0$). The color bar indicates the magnitude of the magnetic field. We see that near the equator, the magnetic fields have an almost negligible poloidal component, with the toroidal component ($B^\phi(r,\theta)$) being dominant; but, as we move away from the equatorial plane, the poloidal components increases and becomes more significant. Additionally, the magnetic field lines are seen to curl around the axis of rotation, indicating that the toroidal structure remains important even at higher latitudes.

By comparing it with the magnetic field around a black hole, we can see that the toroidal components of both remain the same, as expected, since the magnetic field component $B^\phi(r,\theta)$ is independent of the parameter $\lambda$ or the function $R_2(r)$. However, the poloidal component of the black hole's magnetic field is flatter near the equator than that of the wormhole. We also observe that the magnitude of the magnetic field is greater for the black hole than for the wormhole. As we will see in the next section, this correlates with a larger Poynting flux emitted from the black hole.

## 4 Poynting flux

We are now ready to compute the Poynting flux produced in the magnetosphere of the RDSW spacetime and compare the present results with the estimations we previously obtained

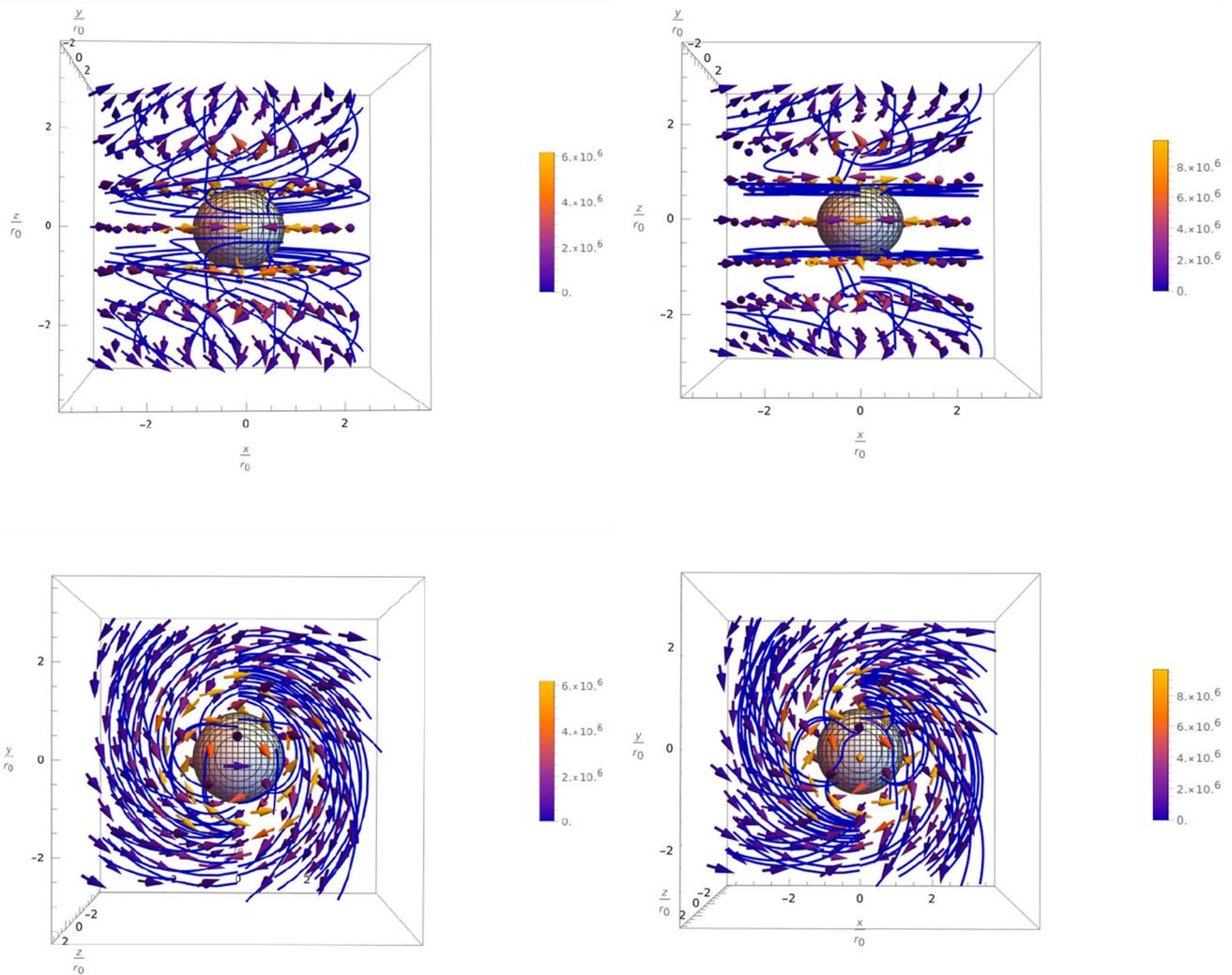

**Fig. 3** Magnetic field lines of a rotating wormhole with $\alpha = 0.5$, $\lambda = \alpha\sqrt{16 + 9\alpha^2}/(4\sqrt{16 - 3\alpha^2})$ (left) next to the ones of a black hole with same $\alpha$ (right), as viewed from the equatorial plane (upper figures) and from the pole (lower figures). $r_0$ is the gravitational radius, $r_0 = 2GM/c^2$

assuming a specific geometry for the magnetic field. Replacing all the expressions in (7), the electromagnetic flux results

$$P_{\text{BZ}} = \frac{2}{c^3} \int_0^{2\pi} d\phi \int_0^{\frac{\pi}{2}} d\theta \sqrt{-g} \mathcal{E}^r(r_+, \theta)$$
$$= -\frac{4\pi}{c^3} \int_0^{\frac{\pi}{2}} d\theta \sqrt{-g} c^2 \omega(r_+, \theta) B^r(r_+, \theta)$$
$$B^\phi(r_+, \theta) \Delta \sin^2 \theta,$$
$$= -\frac{4\pi}{c} \int_0^{\frac{\pi}{2}} d\theta \partial_\theta \left( K \left( (1 - \cos \theta) + \alpha^2 R_2(r_+) \Theta_2(\theta) \right) \right)$$
$$\frac{\alpha i_1(\theta)}{g_T} \alpha \omega_1(\theta) \Delta \sin^2 \theta$$
$$= \frac{4\pi}{c} \int_0^{\frac{\pi}{2}} d\theta \partial_\theta \left( K \left( (1 - \cos \theta) + \alpha^2 R_2(r_+) \Theta_2(\theta) \right) \right)$$
$$\alpha i_1(\theta) \alpha \omega_1(\theta),$$
$$K = \frac{B_0}{c} \left( \frac{GM}{c^2} \right)^3.$$

Substituting $i_1(\theta)$, $\omega_1(\theta)$ and integrating in $\theta$, we obtain that up to second order expansion in the spin parameter, the electromagnetic flux across the throat is

$$P_{\text{BZ}} = \frac{B_0^2 G^2 M^2 \pi \alpha^2}{24 c^3} - \frac{B_0^2 G^2 M^2 \pi R_2(r_+) \alpha^4}{20 c^3}. \quad (49)$$

Since we have employed a perturbative approach in the spin parameter, it is necessary to first determine the range of spin values for which our solution remains valid. For that end, we will compare Eq. (49) in the Kerr limit ($\lambda \to 0$) with an analytical expression for the Poynting flux of a Kerr black hole, given by [2],

$$P_{\text{BZNP}} = \frac{G^2}{c^3} \frac{1 + x^2}{4x^2} \left[ \left( x + \frac{1}{x} \right) \arctan x - 1 \right] B_0^2 M^2 \alpha^2, \quad (50)$$

where $x = a/(cr_+)$. For very low values of the spin parameter on a Kerr black hole, there is a difference between (49) and (50):

$$P_{\text{BZ}}\bigg|_{\lambda=0} = \frac{\pi}{4} P_{\text{BZNP}}, \tag{51}$$

so we will define the corrected Poynting flux,

$$P_{\text{BZ,C}} = \frac{4}{\pi} P_{\text{BZ}}. \tag{52}$$

We plot expressions (50) and (52); the result is shown in Fig. 4. We observe that both expressions coincide for small values of the spin parameter and, as the spin parameter increases, the difference becomes larger and larger. We compute the relative error $(P_{\text{BZP,C}} - P_{\text{BZNP}})/P_{\text{BZNP}}$; this quantity is plotted as a function of the spin parameter in Fig. 5: the blue curve is the relative error between expressions, and the dotted green line represents a relative error of 10%. The latter is reached considering a spin parameter $\alpha \approx 0.5$. This is the upper bound for the spin parameter we consider expression (49) remains valid.

We compute the Poynting flux for three values of the deformation parameter: $\lambda = 0$, $\lambda_{\max}/2$, and $\lambda_{\max}$, where $\lambda_{\max} = \alpha\sqrt{16+9\alpha^2}/(4\sqrt{16-3\alpha^2}) = 0.136743$, and the normalized spin parameter is set to $\alpha = 0.5$. The results are presented in Table 1. We see that as the deformation parameter increases, the flux emitted over the throat decreases, which is a trend we have already observed in [13], which we propose is due to the decrease of the ergo-region dimensions.

Table 1 also includes the values of the Poynting flux $P_{\text{BZ}}^*$ obtained in our previous work [13]. By comparing the new results presented here with those in [13], we observe that the Poynting flux was overestimated for the Kerr black hole and underestimated for the RRDSW when using Wald's model for the magnetic field. This underestimation arose from the fact that, in our previous model, the magnetic field was not defined in certain regions of the ergosphere, leading to the Poynting flux being integrated over a significantly smaller region.

We note, however, that these differences are relatively small; the Poynting fluxes remain of the same order of magnitude in both works. We therefore conclude that, despite the magnetic field in [13] not being derived from the full solution of Maxwell's equations in the RDSW spacetime (as done in the present work) the adopted magnetic field model was sufficiently realistic to capture the main features of the Blandford–Znajek mechanism.

## 5 Conclusions

We have solved Maxwell's equations in the force-free regime to determine the magnetic field geometry in the background spacetime of a rotating wormhole. To the best of our knowledge, this has not been done before. In particular, we have solved the stream equation using the Blandford–Znajek perturbative approach, extending the solution up to second order in the spin parameter.

We found that the first-order solution coincides with the corresponding result in Kerr spacetime. The second-order term, however, increasingly deviates from the Kerr case as the deformation parameter grows. We have also provided visualizations of the magnetic field lines near the wormhole's throat. In the equatorial region, the toroidal component of the magnetic field dominates, while the poloidal component is practically negligible. As one moves toward the poles, the poloidal component becomes more significant. Nonetheless, the magnetic field lines continue to curl around the rotation axis, indicating that the toroidal component remains present throughout.

As mentioned earlier, in a previous work we demonstrated that rotating wormholes can emit an outward electromagnetic flux via the Blandford–Znajek mechanism. In that analysis, the Poynting flux was calculated using a physically motivated magnetic field model, which was not derived as a solution of the stream equation in the corresponding spacetime background. In the present work, we assess the validity of that approximation. We find that the Poynting flux was overestimated for the Kerr black hole and underestimated for the RRDSW when Wald's model was used. Nevertheless, the flux remains of the same order of magnitude in both studies, indicating that the earlier model was sufficiently accurate to capture the leading physical effects.

Although we have determined the geometry of the magnetic field near the throat, it would be interesting to explore how the field behaves through the throat and beyond, on the other side. This is a distinctive feature of wormholes, due to the absence of an event horizon. Addressing this problem requires an appropriate coordinate system that smoothly extends across the throat and covers both regions of the spacetime.

The Blandford–Znajek mechanism should, in principle, be applicable in any modified theory of gravity that admits rotating wormhole solutions with an ergosphere. This, in turn, opens the possibility of comparing the resulting electromagnetic fluxes across different wormhole models, both in general relativity and in alternative theories of gravity. Moreover, since rotating wormhole solutions in modified gravity can circumvent the need for exotic matter—a requirement in general relativity for stable and traversable configurations—we expect the geometry of the magnetic field, and consequently the morphology of the outflows, to differ significantly. We plan to investigate these aspects in future work, as they may provide deeper insights into the global structure of magnetospheres in wormhole geometries.

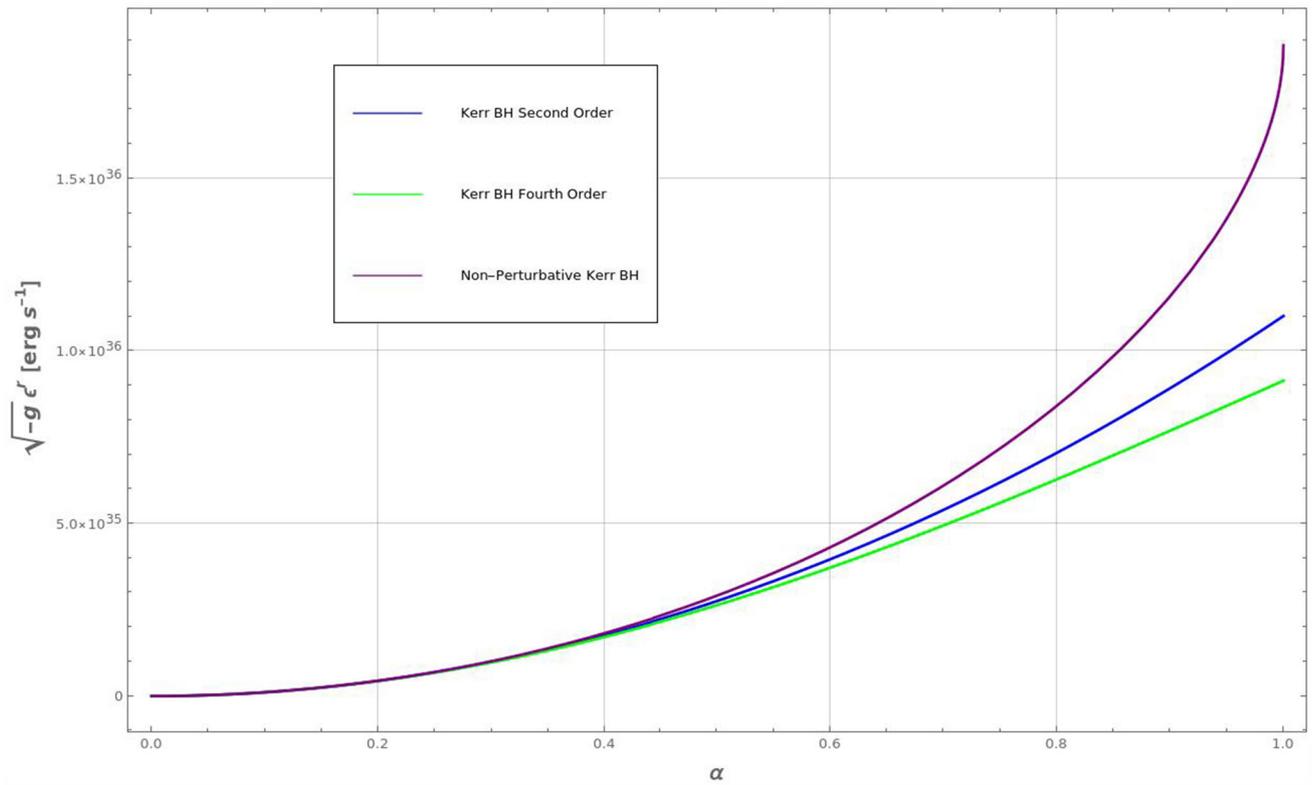

**Fig. 4** Perturbative and non-perturbative BZ Poynting Flux for a Kerr Black Hole

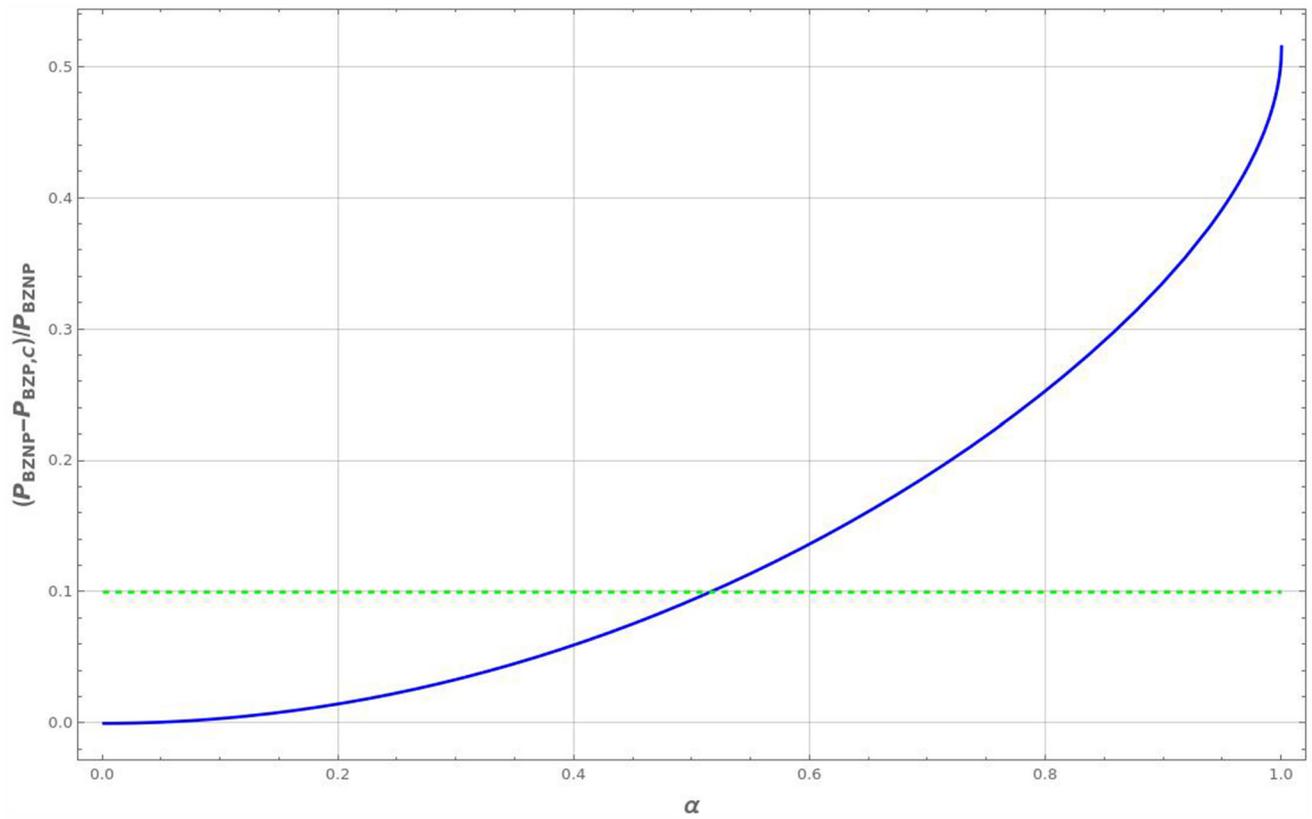

**Fig. 5** Relative error of the Perturbative BZ vs. Expression (50). The green dashed line represents an error of 10%

**Table 1** Maximum energy extraction rates for $a = 0.5\,GM/c^2$ and different deformation parameter $\lambda$. $P_{\text{BZNP}}$ is given by expression (50), $P_{\text{BZ}}$ are this work results and $P_{\text{BZ}}^*$ are results derived by the model used in [13]

| $\alpha = 0.5$ | $P_{\text{BZNP}}\,[10^{35}\,erg\,s^{-1}]$ | $P_{\text{BZ,C}}\,[10^{35}\,erg\,s^{-1}]$ | $P_{\text{BZ}}^*\,[10^{35}\,erg\,s^{-1}]$ |
| --- | --- | --- | --- |
| $\lambda=0$ | 2.91016 | 2.63636 | 5.82706 |
| $\lambda=\frac{\lambda_{max}}{2}$ | – | 2.61508 | 0.971967 |
| $\lambda=\lambda_{max}$ | – | 2.59373 | 0.383509 |

## 6 Appendix: Magnetic field lines in different coordinate systems

In what follows, we show how to transform the magnetic field components in Boyer-Lindquist coordinates to Cartesian coordinates, that is

$$B_r(r,\theta) \longrightarrow B_x(x,y,z),$$
$$B_\theta(r,\theta) \longrightarrow B_y(x,y,z),$$
$$B_\phi(r,\theta) \longrightarrow B_z(x,y,z). \quad (53)$$

To do this, we must transform the coordinates $r$, $\theta$, $\phi$ into the coordinates $x$, $y$, $z$ and the vectorial basis $\hat{e}_r$, $\hat{e}_\theta$, $\hat{e}_\phi$ in the vector basis $\hat{e}_x$, $\hat{e}_y$, $\hat{e}_z$, which is described as

$$r(x,y,z) = \frac{\sqrt{-a^2 + x^2 + y^2 + z^2 + \sqrt{4a^2 z^2 + (a^2 - x^2 - y^2 - z^2)^2}}}{\sqrt{2}},$$
$$\theta(x,y,z) = \arccos\left(\frac{z}{r}\right),$$
$$\phi(x,y,z) = \arctan\left(\frac{y}{x}\right), \quad (54)$$

and

$$\hat{e}_r(r,\theta) = \frac{r}{\sqrt{r^2 + a^2 \cos^2\theta}}(\sin\theta\cos\phi\,\hat{e}_x + \sin\theta\sin\phi\,\hat{e}_y)$$
$$+ \sqrt{\frac{r^2 + a^2}{r^2 + a^2 \cos^2\theta}}\cos\theta\,\hat{e}_z,$$
$$\hat{e}_\theta(r,\theta) = \sqrt{\frac{r^2 + a^2}{r^2 + a^2 \cos^2\theta}}(\cos\theta\cos\phi\,\hat{e}_x + \cos\theta\sin\phi\,\hat{e}_y)$$
$$- \frac{r}{\sqrt{r^2 + a^2 \cos^2\theta}}\sin\theta\,\hat{e}_z,$$
$$\hat{e}_\phi(r,\theta) = -\sin\theta\,\hat{e}_x + \cos\theta\,\hat{e}_y. \quad (55)$$

Thus, the Cartesian components of the magnetic field are as follows

$$B_x(x,y,z) = B_r(r,\theta)\hat{e}_{r,x}(r,\theta,\phi)$$
$$+ B_\theta(r,\theta)\hat{e}_{\theta,x}(r,\theta,\phi) + B_\phi(r,\theta)\hat{e}_{\phi,x}(r,\theta,\phi),$$
$$B_y(x,y,z) = B_r(r,\theta)\hat{e}_{r,y}(r,\theta,\phi) + B_\theta(r,\theta)\hat{e}_{\theta,y}(r,\theta,\phi)$$
$$+ B_\phi(r,\theta)\hat{e}_{\phi,y}(r,\theta,\phi),$$
$$B_z(x,y,z) = B_r(r,\theta)\hat{e}_{r,z}(r,\theta,\phi) + B_\theta(r,\theta)\hat{e}_{\theta,z}(r,\theta,\phi)$$
$$+ B_\phi(r,\theta)\hat{e}_{\phi,z}(r,\theta,\phi). \quad (56)$$

where $\hat{e}_{\gamma,m}$ is the $m$ Cartesian component of the $\gamma$ component of the Boyer-Lindquist vector basis.

**Acknowledgements** D. P. acknowledges the support from CONICET under Grant No. PIP 0554 and AGENCIA I+D+i under Grant PICT-2021-I-INVI-00387.

**Author contributions** All authors contributed equally to the manuscript.

**Funding** D. P. acknowledges the support from CONICET under Grant No. PIP 0554 and AGENCIA I+D+i under Grant PICT-2021-I-INVI-00387.

**Data Availability Statement** This manuscript has no associated data. [Author's' comment: Data sharing not applicable to this article as no datasets were generated or analysed during the current study.].

**Code Availability Statement** This manuscript has no associated code/software. [Author's comment: Code/Software sharing not applicable to this article as no code/software was generated or analysed during the current study.].

**Declarations**

**Conflict of interest** Not declared.

**Ethics approval** Not declared.

**Consent to participate** Not declared.

**Consent for publication** Not declared.

**Availability of data and materials** Not declared.